\documentstyle[prl,aps,epsf]{revtex}
\begin{document}
\title{Quantum entropy of a non-extreme  stationary axisymmetric \\ black hole
 due to minimally coupled quantum scalar field$^*$\footnotetext[1]{
 This work was
partially supported by the National Natural Science Foundation of
China and Natural Science Foundation of Hunan Province}}
\author{Jiliang Jing $^{a b }$  \ \ \ \ Mu-Lin Yan $^{b}$}
\address{a) Physics Department and Institute of Physics , Hunan Normal
University,\\ Changsha, Hunan 410081, P. R. China;  \\ b)
Department of Astronomy and Applied Physics, University of Science
and Technology of China, \\ Hefei, Anhui 230026, P. R. China}
\maketitle
\begin{abstract}

By using the 't Hooft ``brick wall" model and the Pauli-Villars
regularization scheme we calculate the statistical-mechanical
entropy arising from the minimally coupled scalar fields which
rotate with the azimuthal angular velocity $\Omega_0=\Omega_H$
($\Omega_H$ is the angular velocity of the black hole horizon) in
the general four-dimensional non-extreme  stationary axisymmetric
black hole space-time. We also show, for the Kerr-Newman and the
Einstein-Maxwell dilaton-axion black holes, that the
statistical-mechanical entropy obtained from our derivation and
the quantum thermodynamical entropy by the conical singularity
method are equivalent.

\end{abstract}

\section{INTRODUCTION}
\label{sec:intro} \vspace*{0.2cm}

Since Bekenstein and Hawking found that the black hole entropy is
proportional to the event horizon area by comparing black hole
physics with thermodynamics and from the discovery of the black
hole evaporation \cite{Bekenstein72}-\cite{Bekenstein74}, many
efforts have been devoted to study the statistical origin of the
black hole entropy. Especially, the idea to relate the entropy of
the black hole to quantum excitations of the black hole has
attracted many attentions \cite{Hooft85} -\cite{Mann96}. The
thermodynamical entropy of the black hole is related to the
covariant Euclidean free energy $F^E[g, \beta]=\beta^{-1}W[g,
\beta]$\cite{Frolov98}, where $\beta$ is the inverse temperature.
The function $W[g, \beta]$ is given on Euclidean manifolds with
the period $\beta$ in the Euclidean time $\tau$. We can calculate
the free energy $F^E$ by the conical singularities method. This
procedure was consistently carried out for the studies of the
static black holes and the rotating charged Kerr black
hole\cite{Solodukhin951} \cite{Solodukhin952}  \cite{Fursaev95}
\cite{Fursaev96} \cite{Mann96}. On the other hand, the canonical
statistical-mechanical entropy can be derived from the free energy
$F^C$ of a system\cite{Frolov98}, where $F^C$ can be defined in
term of the one particle spectrum. One of the ways to calculate
$F^C$ is the ``brick wall" model (BWM) proposed by 't Hooft
\cite{Hooft85}. He argued that the black hole entropy is
identified with the statistical-mechanical entropy arising from a
thermal bath of quantum fields propagating outside the horizon. In
this model, in order to eliminate divergence which appears due to
the infinite growth of the density of states closed to the
horizon, 't Hooft introduces a ``brick wall" cutoff: a fixed
boundary $\Sigma_h$ near the event horizon within the quantum
field does not propagate and the Dirichlet boundary condition was
imposed on the boundary, i. e., wave function $\phi=0$ for
$r=r({\Sigma_h})$. Later, J. G. Demers, R, Lafrance and R. C.
Myers \cite{Demers95} pointed out that the Dirichlet condition can
be removed if we use the Pauli-Villars regulated theory. The BWM
has been successfully used in studies of the
statistical-mechanical entropy for many black holes
\cite{Hooft85}, \cite{Ghosh94}-\cite{Ho98}, \cite{Belgiorno96},
\cite{Jing97}.

Recently, Frolov and Fursaev\cite{Frolov98} reviewed the  studies
of the relation between the thermodynamic entropy and the
statistical-mechanical entropy of  the  black holes. They shown
that  for the general static black holes the covariant Euclidean
free energy $F^E$ and the statistical-mechanical free energy $F^C$
are equivalent when ones use the ultraviolet regularization method
\cite{Frolov98}.

As static case, the quantum entropy for the stationary
axisymmetric black holes has also been studied by many authors
recently. Mann and Solodukhin\cite{Mann96} investigated the
covariant Euclidean formulation for the Kerr-Newman black hole.
They showed that an Euclidean manifold which was obtained by Wick
rotation of the Kerr-Newman geometry with Killing horizon has a
conical singularity similar to the one which appears in the static
black holes. The one-loop quantum correction to the entropy of the
charged Kerr black hole was calculated by applying the method of
the conical singularities. They found an interesting result that
the logarithmic term of the quantum entropy for the Kerr-Newman
black hole can be written as a constant plus a term proportional
to the charge and so, for the Schwarzschild and the Kerr black
holes, the logarithmic parts in the entropy  are exactly equal.

Cognola \cite{Cognola98}, through the Euclidean path integral and
using heat kernel and $\zeta$-function regularization scheme,
studied the one-loop contribution to the entropy for a scalar
field in the Kerr black hole. In the calculation he took an
approximation of the metric, which, after a conformal
transformation, takes a Rindler-like form. He pointed out in
Ref.\cite{Cognola98} that the result is valid also for the
Kerr-Newman black hole. Nevertheless, the result is in contrast
with the corresponding one obtained in Ref.\cite{Mann96}.

In Refs. \cite{Lee96} \cite{Lee961} \cite{Ho96},  by using the 't
Hooft BWM Lee and Kim, Ho, Kim and Park  discussed the
statistical-mechanical entropy of some stationary black holes,
such as Kerr black hole, Kerr-Newman black hole, and Kaluza-Klein
black hole. The results showed that the entropies can be expressed
as $k\frac{A_H}{\varepsilon ^2}$, where $A_H$ is the area of the
event horizon, and $\varepsilon=\int^{r_H+h}_{r_H}dr\sqrt{g_{rr}}$
is the proper distance from the horizon to the $r_H+h$ and the $h$
the cutoff in the radial coordinate near the horizon. However, the
logarithmically divergent term of the statistical-mechanical
entropy for the four-dimensional stationary axisymmetric black
hole space-time was not be investigated.

Although much attention has been paid on the study of the quantum
entropy of the stationary axisymmetric black holes, the relation
between the statistical-mechanical entropy and the thermodynamical
entropy  for the rotating black holes has not been investigated
yet\cite{Frolov98}. The aim of this paper is to obtained an
expression of the general statistical-mechanical  entropy for the
general four-dimensional non-extreme  stationary axisymmetric
black hole by using the BWM and the Pauli-Villars regularization
scheme, and then make a comparison of the statistical-mechanical
entropy obtained by using BWM and the thermodynamical entropy by
the conical singularity method for some well-known stationary
axisymmetric black holes.

The  paper is organized  as follows: In sec.2, the general
stationary axisymmetric black hole is introduced and some
properties of the  black hole that it is necessary to understand
the thermodynamics of quantum field  are studied.  In sec. 3,
making use of 't Hooft's BWM \cite{Solodukhin97} we  deduce  a
formula of the statistical-mechanical entropy  for the general
non-extreme  stationary axisymmetric black hole. In the last
section, the statistical-mechanical entropies for the Kerr-Newman
and the Einstein-Maxwell dilaton-axion (EMDA) black holes  are
studied by using the formula. Then the results are compared with
the entropy obtained by the conical singularity method. Finally,
we end with some conclusions.

\section{the space-time of the general  non-extreme  \\
         Stationary axisymmetric black hole}
\vspace*{0.4cm}

In Boyer-Lindquist coordinates the most general line element for a
stationary axisymmetric black hole in four-dimensional space-time
can be expressed as
\begin{equation}
ds^2=g_{tt}dt^2+g_{rr}dr^2+g_{t \varphi}dtd\varphi+g_{\theta
\theta}d\theta ^2+g_{\varphi \varphi}d\varphi ^2,
 \label{metric0} \end{equation}
where $g_{tt}$, $g_{rr}$, $g_{t\varphi}$, $g_{\theta \theta}$ and
$g_{\varphi \varphi}$ are functions of the coordinates $r$ and
$\theta$ only. Because the   space-time (\ref{metric0}) is
stationary and axisymmetric one it  exists a stationary  Killing
vector field $\xi^{\mu}=(1, 0, 0, 0)$ and an axial Killing field
$\Psi^{\mu}=(0,0,0,1)$\cite{Carmeli}. By taking a liner
combination of  $\xi^\mu$ and $\Psi^\mu$  we obtain a new Killing
field
\begin{equation}
l^\mu=\xi^\mu+\Omega_{H}\Psi^\mu,\label{lll}
\end{equation}
which is normal to the  horizon of the black hole. In Eq.
(\ref{lll}) the constant $\Omega _{H}$ is called the angular
velocity of the event horizon. An interesting feature of the
 black hole worthy of note is that the
norm of the Killing field $l^\mu$ is zero on the horizon since the
horizon is a null surface and vector $\l^\mu$ is normal to the
horizon. That is to say, the black hole horizon is a surface where
the Killing field $l^\mu$ is null. Substituting $l^\mu$ into the
formula of the surface gravity \cite{Wald} $
\kappa^2=-\frac{1}{2}\l^{\mu
;\nu}l_{\mu ;\nu}, $ we obtain
\begin{eqnarray}
\kappa&=&\frac{-1}{2}\lim_{r\rightarrow r_H}\left
(\sqrt{\frac{-1}{g_{rr}\left(g_{tt}-
\frac{g_{t\varphi}^2}{g_{\varphi \varphi}}\right)}}\frac{d}{dr}
\left(g_{tt}-\frac{g_{t\varphi}^2}{g_{\varphi \varphi}}\right)
\right)\nonumber \\
&=&\frac{2\pi}{\beta_H}
, \label{surface}
\end{eqnarray}
where $r_H$ represents the outermost event horizon, $1/\beta_H$ is
the Hawking temperature, and here and hereafter the metric
signature is taken as $(-, +, +, +)$. We know that the event
horizon is a null hypersurface determined by
\begin{equation}
g^{\mu \nu}\frac{\partial H}{\partial x^\mu}\frac{\partial H}{\partial
x^\nu}=0. \label{hs1}
\end{equation}
For the stationary axisymmetric black hole (\ref{metric0})  the
function  $H$ is a function of  $r$ and $\theta$ only.
Substituting the  metric (\ref{metric0}) into Eq. (\ref{hs1}) and
discussing carefully we find
 \begin{equation}
\frac{1}{g^{tt}(r_H)}=\left(g_{tt}-\frac{g_{t\varphi}^2}{g_{\varphi\varphi}}
\right)_{r_H}=0. \label{nullcon}
\end{equation}
Solutions of which determine the location of  the event horizons.
From Eq. (\ref{nullcon}) we know that  for a non-extreme
stationary axisymmetric black hole $1/g^{tt}$ can be expressed as
 \begin{equation}
\left(g_{tt}-\frac{g_{t\varphi}^2}{g_{\varphi\varphi}}
\right)\equiv G_1(r,\theta)(r-r_H), \label{asum}
\end{equation}
where $G_1(r,\theta)$ is a regular function in the region $ \infty>r\geq
r_H$ and its value is nonzero on the outermost event  horizon
$ r_H$. On the other hand, since $\kappa=constant$
 and $\frac{1}{g^{tt}}=0$ on the event horizon $r=r_H$, we find from
 Eq.(\ref{surface}) that $g^{rr}$ must   take the following form
\begin{equation}
g^{rr}\equiv G_2(r,\theta)(r-r_H),\label{asum1}
\end{equation}
where $G_2(r,\theta)$ is a well-define function in the region $
\infty>r\geq r_H$ and is nonzero on the horizon $r_H$ too. Making
use of Eqs. (\ref{asum}) and (\ref{asum1}), we obtain
 \begin{equation}
g_{rr}\left(g_{tt}-\frac{g_{t\varphi}^2}{g_{\varphi\varphi}}
\right)=\frac{G_1(r, \theta)}{G_2(r, \theta)}
\equiv -f(r,\theta), \label{relation}
\end{equation}
where the $f(r, \theta)$ is a constant or a regular  function on the
outermost event horizon and outside the horizon.

\section{The statistical-mechanical entropy  of general \\
 non-extreme  stationary axisymmetric black hole} \vspace*{0.5cm}

We now try to find an expression of the statistical-mechanical
entropy due to the minimally coupled quantum scalar fields in a
general four-dimensional stationary axisymmetric black hole.  We
first seek  the total number of modes with energy less than $E$ by
using Klein-Gordon equation, and then calculate a free energy. The
statistical-mechanical entropy of the black hole is obtained by
the variation of the free energy with respect to inverse
temperature and setting $\beta =\beta_H$.

Using WKB approximation with
\begin{equation}
\phi =exp[-iEt+im\varphi+iW(r,\theta)], \label{phi}
\end{equation}
and substituting  the metric (\ref{metric0}) into  the
Klein-Gordon equation of the scalar field $\phi$ with mass $\mu$
and nonminimal $\xi R \phi^2$ ($R$ is the scalar curvature of the
spacetime) coupling
\begin{equation}
\frac{1}{\sqrt{-g}}\partial _\mu(\sqrt{-g}g^{\mu\nu}\partial _\nu
\phi)-(\mu ^2+\xi R)\phi=0, \label{kg}
\end{equation}
we find \cite{Mann92}
\begin{equation}
p_r^2=\frac{1}{g^{rr}}
[-g^{tt}E^2+2g^{t\varphi}Em-g^{\varphi\varphi}
m^2-g^{\theta\theta}p_\theta^2-(\mu ^2+\xi R)],\label{W}
\end{equation}
 where $p_r\equiv \partial _r W(r, \theta)$ and  $p_\theta \equiv
 \partial_\theta W(r, \theta)$.
If the scalar curvature $R$ takes a nonzero value at the horizon
then this region can be approximated by some sort of constant
curvature space. However, the exact results for such a black hole
showed that the mass parameter in the solution enters only in the
combination $(\mu^2-R/6)$\cite{Solodukhin97}\cite{Birrell82}.
Therefore, inserting the covariant metric into Eq.(\ref{W}) we
arrive at
\begin{equation}
p_r^2=-\frac{g_{rr}g_{\varphi\varphi}}
{g_{tt}g_{\varphi\varphi}-g_{t\varphi}^2} \left[(E-\Omega
m)^2+\left(g_{tt}-\frac{g_{t\varphi}^2}{g_{\varphi\varphi}}
\right)\left(\frac{m^2}{g_{\varphi\varphi}}+\frac{p_\theta^2}
{g_{\theta\theta}}+M^2(r,\theta)\right)\right],\label{pr}
\end{equation}
where $\Omega \equiv -\frac{g_{t \varphi }}{g_{\varphi \varphi}}$
and $M^2(r,\theta)\equiv \mu^2-(\frac{1}{6}-\xi)R$. In this paper
our discussion is restricted to study minimally coupled ($\xi=0$)
scalar fields. We know from Eq. (\ref{pr}) that $W(r,\theta)$ can
be expressed as
 \begin{equation}
W(r,\theta)=\pm \int^r
\sqrt{\frac{-g_{rr}g_{\varphi\varphi}}{g_{tt}g_{\varphi\varphi}-
g_{t\varphi }^2}} K(r, \theta) dr+c(\theta),
\end{equation}
where
 \begin{equation}
K(r, \theta)=\sqrt{(E-\Omega m)^2+\left(g_{tt}-
\frac{g_{t\varphi}^2}{g_{\varphi\varphi}}
\right)\left(\frac{m^2}{g_{\varphi\varphi}}+\frac{p_\theta^2}
{g_{\theta\theta}}+M^2(r, \theta)\right)}.\label{KK}
\end{equation}
Therefore, in phase space the  number of the modes with $E$, $m$
and $p_{\theta}$ is shown by\cite{Padmanabhan86}
\begin{equation}
n(E, m, p_{\theta})=\frac{1}{\pi} \int d\theta
\int^{r_E}_{r_H+h}\sqrt{\frac{-g_{rr}g_{\varphi
\varphi}}{g_{tt}g_{\varphi \varphi}-g_{t\varphi}^2}}K(r,\theta)dr.
\end{equation}

Zhao and Gui \cite{Zhao83} pointed out that ``a physical space"
must be dragged by the gravitational field with a azimuth angular
velocity $\Omega_{H}$ in the stationary axisymmetric space-time
(\ref{metric0}). Apparently, a  quantum scalar field in thermal
equilibrium at temperature $1/\beta$ in the stationary
axisymmetric black hole must be dragged too. Therefore, it is
rational to assume that the scalar field is rotating with angular
velocity $\Omega_0=\Omega_{H} $. For such an equilibrium ensemble
of states of the scalar field, the free  energy is given by
\begin{eqnarray}
\beta F&=& \int dm \int dp_{\theta}\int dn(E, m, p_{\theta})ln
\left[ 1-e^{-\beta (E-\Omega_0 m)}\right] \nonumber \\ &=&\int dm
\int dp_{\theta}\int dn(E+\Omega_0 m, m, p_{\theta})ln \left(
1-e^{-\beta E}\right) \nonumber \\ &=&-\beta \int dm \int
dp_{\theta}\int \frac{n(E+\Omega_0 m, m, p_{\theta})} {e^{\beta
E}-1} dE \nonumber \\ &=&-\beta \int \frac{n(E)}{e^{\beta E}-1} dE
, \label{f1}
\end{eqnarray}
with
\begin{eqnarray}
n(E)&=&\int dm \int dp_{\theta}\int n(E+\Omega_0 m, m, p_{\theta})
\nonumber \\ &=&\frac{1}{3\pi}\int d\theta
\int^{r_E}_{r_H+h}\frac{dr \sqrt{g_4} }
{\left[\left(g_{tt}-\frac{g_{t\varphi }^2}{g_{\varphi
\varphi}}\right)\left(1+\frac{g_{\varphi
\varphi^2(\Omega-\Omega_0)^2}}{g_{tt}g_{\varphi \varphi}-g_{t
\varphi}^2} \right)\right]^2} \left[E^2 \right. \nonumber \\ & &
\left. +\left(g_{tt}-\frac{g_{t\varphi}^2}{g_{\varphi \varphi}}
\right)\left(1+\frac{g_{\varphi
\varphi^2(\Omega-\Omega_0)^2}}{g_{tt}g_{\varphi \varphi}-g_{t
\varphi}^2} \right)M^2(r,\theta)\right]^{\frac{3}{2}}. \label{nE}
\end{eqnarray}
where the function $n(E)$ presents the total number of the modes
with energy less than $E$. The integrations of the $m$ and
$p_{\theta}$ in the Eq.(\ref{f1}) are taken only over the value
for which the square root in Eq.(\ref{KK}) exists.

Taking the integration of the $r$ in the Eq.(\ref{nE}) for the
case $\Omega_0=\Omega_H$ we have
\begin{eqnarray}
n(E)&=&-\frac{1}{2\pi}\int d\theta \left \{\sqrt{g_{\theta
\theta}g_{\varphi \varphi}}\left[\frac{2}{3}\left(\frac{E
\beta_H}{4\pi} \right)^3C(r,\theta)+M^2(r,\theta) \left(\frac{E
\beta_H}{4\pi}\right)\right] ln \left(\frac{E^2}{E^2_{min}}\right)
\right\}_{r_H}
 \nonumber \\
& &-\frac{1}{3\pi}\left(\frac{\beta_H}{4\pi}\right)\int d\theta
\left\{\sqrt{g_{\theta
\theta}g_{\varphi\varphi}}M^2(r,\theta)
\left(E-\frac{E^3}{E^2_{min}}\right)\right\}_{r_H}
, \label{n0}
\end{eqnarray}
where
\begin{eqnarray}
C(r,\theta)&=&\frac{\partial ^2g^{rr}}{\partial
r^2}+\frac{3}{2}\frac{\partial g^{rr}}{\partial r}\frac{\partial
\ln f}{\partial
r}-\frac{2\pi}{\beta_H\sqrt{f}}\left(\frac{1}{g_{\theta
\theta}}\frac{\partial g_{\theta \theta}}{\partial r}+
\frac{1}{g_{\varphi \varphi}}\frac{\partial g_{\varphi
\varphi}}{\partial r}\right)-\frac{2g_{\varphi
\varphi}}{f}\left[\frac{\partial}{\partial r}\left(\frac{g_{t
\varphi}}{g_{\varphi \varphi}}\right)\right]^2,\nonumber\\
 E^2_{min}&=&-M^2(r_H, \theta)\left(g_{tt}-\frac{g_{t\varphi^2}}{g_{\varphi
\varphi}}\right)_{\Sigma_h}, \nonumber \\
\bar{K}^2&=&E^2+\left(g_{tt}-\frac{g_{t\varphi}^2}{g_{\varphi
\varphi}}\right)_{\Sigma_h}M^2(r_H, \theta),
\end{eqnarray}
here and hereafter $f\equiv f(r,\theta)$ which is defined by Eq.
(\ref{relation}).

Now let us use the Pauli-Villars regularization scheme
\cite{Demers95} by introducing five regulator fields $\{\phi_i,
i=1,...,5\}$ of different statistics with masses
$\{\mu_i,i=1,...,5\}$ dependent on the UV cutoff \cite{Demers95}.
If we rewrite the original scalar field $\phi=\phi_0$ and
$\mu=\mu_0$, then these fields satisfy $\Sigma^5_{i=0}\triangle
_i=0$ and $\Sigma^5_{i=0}\triangle _i \mu^2_i=0$, where
$\triangle_0=\triangle_3=\triangle_4=+1$ for the commuting fields
and $\triangle_1=\triangle_2=\triangle_5=-1$ for the anticommuting
fields. Since each of the fields makes a contribution to the free
energy of Eq.(\ref{f1}), the total free energy can be expressed as
\begin{eqnarray}
\beta \bar{F}=\sum^5_{i=0}\triangle _i \beta F_i. \label{f2}
\end{eqnarray}
Substituting Eq.(\ref{f1}) into (\ref{n0}) and then taking the
integration over E we find
\begin{eqnarray}
\bar{F}&=&\frac{-1}{48}\frac{\beta_H}{\beta ^2}\int
d\theta\left\{\sqrt{g_{\theta \theta}g_{\varphi
\varphi}}\right\}_{r_H} \sum^5_{i=0}\triangle _i
M_i^2(r_H,\theta)lnM_i^2(r_H, \theta)
-\frac{1}{2880}\frac{\beta_H^3}{\beta^4} \int
d\theta\left\{\sqrt{g_{\theta \theta}g_{\varphi \varphi}} \right.
\nonumber \\ & &\times \left. \left[\frac{\partial
^2g^{rr}}{\partial r^2}+ \frac{3}{2}\frac{\partial
g^{rr}}{\partial r}\frac{\partial \ln f}{\partial
r}-\frac{2\pi}{\beta_H \sqrt{f}}\left(\frac{1}{g_{\theta
\theta}}\frac{\partial g_{\theta \theta}}{\partial r}+
\frac{1}{g_{\varphi \varphi}}\frac{\partial g_{\varphi
\varphi}}{\partial r}\right) -\frac{2g_{\varphi
\varphi}}{f}\left[\frac{\partial}{\partial r}\left(\frac{g_{t
\varphi}}{g_{\varphi \varphi}}\right)\right]^2
\right]\right\}_{r_H} \nonumber \\ & &\times
 \sum^5_{i=0}\triangle
_ilnM_i^2(r_H,\theta),  \label{f-0}
\end{eqnarray}
where $M_i^2(r_H, \theta)=\mu_i^2-\frac{1}{6}R$. Then the  total
statistical-mechanical entropy at the Hawking temperature
$\frac{1}{\beta}=\frac{1}{\beta_H}$ is given by
\begin{eqnarray}
S&=&\left[\beta^2 \frac{\partial \bar{F}}{\partial
\beta}\right]_{\beta=\beta_H}\nonumber \\ &=&\frac{1}{48\pi}\int
 d\theta d\varphi\left(\sqrt{g_{\theta \theta}g_{\varphi
\varphi}}\right)_{r_H} \sum^5_{i=0}\triangle _i
M_i^2(r_H,\theta)lnM_i^2(r_H, \theta) +\left\{\frac{1}{32\times
45\pi}\int d\theta d\varphi \sqrt{g_{\theta \theta}g_{\varphi
\varphi}}  \right. \nonumber \\ & & \left. \times
\left[\frac{\partial ^2g^{rr}}{\partial r^2} + \frac{ 3 }{ 2
}\frac{ \partial g^{rr} }{
\partial r } \frac{\partial \ln f}{\partial r}-\frac{2\pi}{\beta_H
\sqrt{f}}\left(\frac{1}{g_{\theta \theta}}\frac{\partial g_{\theta
\theta}}{\partial r}+ \frac{1}{g_{\varphi \varphi}}\frac{\partial
g_{\varphi \varphi}}{\partial r}\right)-\frac{2 g_{\varphi
\varphi}}{f}\left[\frac{\partial}{\partial r}\left(\frac{g_{t
\varphi}}{g_{\varphi
\varphi}}\right)\right]^2\right]\right\}_{r_H}\nonumber \\ &
&\times
 \sum^5_{i=0}\triangle _ilnM_i^2(r_H,
\theta). \label{SM}
\end{eqnarray}
Using  the assumption that the scalar curvature $R$ at the horizon
is much smaller than each $\mu_i$ and noting that the area of the
event horizon is given by $A_{\Sigma}= \int d\varphi \int
d\theta\left\{\sqrt{g_{\theta \theta}g_{\varphi
\varphi}}\right\}_{r_H}$, we obtain at last the following
expression of the statistical-mechanical entropy
\begin{eqnarray}
S&=&\frac{A_{\Sigma}}{48\pi}\sum^5_{i=0} \triangle _i
\mu_i^2ln\mu_i^2+\left\{-\frac{1}{6\times 48 \pi} \int d\theta
d\varphi \left(R\sqrt{g_{\theta \theta}g_{\varphi
\varphi}}\right)_{r_H}+\frac{1}{32\times 45\pi}\int d\theta
d\varphi \right.\nonumber \\ & & \times \left.
\left[\sqrt{g_{\theta \theta}g_{\varphi \varphi}}\left(
\frac{\partial ^2g^{rr}}{\partial r^2}+ \frac{3}{2}\frac{\partial
g^{rr}}{\partial r}\frac{\partial \ln f}{\partial
r}-\frac{2\pi}{\beta_H \sqrt{f}}\left(\frac{1} {g_{\theta
\theta}}\frac{\partial g_{\theta \theta}}{\partial r}+
\frac{1}{g_{\varphi \varphi}}\frac{\partial g_{\varphi
\varphi}}{\partial r}\right) -\frac{2g_{\varphi \varphi}}{f}
\right. \right. \right. \nonumber \\ & & \left.  \left. \left.
\times \left[\frac{\partial}{\partial r}\left(\frac{g_{t
\varphi}}{g_{\varphi
\varphi}}\right)\right]^2\right)\right]_{r_H}\right\}
\sum^5_{i=0}\triangle _iln\mu_i^2. \label{smu}
\end{eqnarray}
which is valid for the general non-extreme  stationary
axisymmetric black holes which the metric can be expressed as
(\ref{metric0}) in the Boyer-Lindquist coordinates and their
signature is  $(-, +, +, +)$. For the black holes with signature
$(+, -, -, -)$ a corresponding formula can be obtained by replaced
the $\beta_H$ with $-\beta_H$ in the Eq.(\ref{smu}).

\section{discussion and summary}

In this section, let begin discussion with the study of the
statistical-mechanical entropy of the Kerr-Newman black hole and
EMDA black hole by using the formula (\ref{smu}).

\vspace*{0.4cm}
 {\bf (A) The entropy of the Kerr-Newman black
hole}
\vspace*{0.4cm}

In Boyer-Lindquist coordinates, the metric of the Kerr-Newman
black hole \cite{Kerr63}\cite{Newman65} takes the form
\begin{eqnarray}
g_{tt}&=&-\frac{\bigtriangleup -a^2sin^2\theta }{\rho ^2}, \ \ \ \
 g_{t\varphi}=-\frac{asin^2\theta
(r^2+a^2-\bigtriangleup)}{\rho ^2},\nonumber \\
g_{rr}&=&\frac{\rho ^2}{\bigtriangleup}, \ \ \ \ g_{\theta \theta
}=\rho ^2, \ \ \ \ g_{\varphi \varphi
}=\left(\frac{(r^2+a^2)^2-\bigtriangleup a^2sin^2\theta }{\rho
^2}\right)sin^2 \theta,\label{knm}
\end{eqnarray}
with
\begin{equation}
\rho^2=r^2+a^2cos^2\theta,\ \ \ \ \bigtriangleup=(r-r_+)(r-r_-),
\end{equation}
where $r_+=r_H=M+\sqrt{M^2-Q^2-a^2}$, $r_-=M-\sqrt{M^2-Q^2-a^2}$,
and $M$,  $Q$ represent the mass and charge of the black hole,
respectively.  Using the metric (\ref{knm}) we get
\begin{eqnarray}
& &\left\{\frac{\partial ^2g^{rr}}{\partial r^2}+
\frac{3}{2}\frac{\partial g^{rr}}{\partial r}\frac{\partial \ln
f}{\partial r}-\frac{2\pi}{\beta_H \sqrt{f}}\left(\frac{1}
{g_{\theta \theta}}\frac{\partial g_{\theta \theta}}{\partial r}+
\frac{1}{g_{\varphi \varphi}}\frac{\partial g_{\varphi
\varphi}}{\partial r}\right) -\frac{2g_{\varphi \varphi}}{f}
\left[\frac{\partial}{\partial r}\left(\frac{g_{t
\varphi}}{g_{\varphi
\varphi}}\right)\right]^2\right\}_{r_+}\nonumber
\\ &=&\frac{16r_+^2[(r_+r_--a^2)-(r_++r_-)r_+]+4[(a^2-r_+r_-)+3(r_
++r_-)r_+]\rho^2}{\rho^6}+\frac{2(a^2-r_+r_-)}{\rho^4}\nonumber
\\ & &+\frac{2a^2(1+cos^2\theta )\rho ^2-8a^2(r_+^2+a^2)cos^2\theta
}{\rho^6}, \nonumber \\  R&=&0. \label{kn1}
\end{eqnarray}
Inserting Eq.(\ref{kn1}) into Eq.(\ref{smu}) and then taking the
integrations of the $\theta$ and $\varphi$ we find that the
statistical-mechanical entropy of the Kerr-Newman black hole is
given by
\begin{eqnarray}
S_{KN}&=&\frac{A_{\Sigma}}{48\pi}\sum^5_{i=0} \triangle _i
\mu_i^2ln\mu_i^2-\frac{1}{90}\left\{1+\frac{3(a^2-r_+
r_-)}{4r_+^2}\left[1+\frac{r_+^2+a^2}{ar_+}Arctan\left(
\frac{a}{r_+}\right)\right]\right\} \sum^5_{i=0}\triangle
_iln\mu_i^2, \nonumber\\ \label{kn2}
\end{eqnarray}
where $A_{\Sigma}=4\pi(r_+^2+a^2)$. Noting $r_+r_--a^2=Q^2$ and
the Pauli-Villars regularization scheme caused a factor
$-\frac{1}{2}$ for the second part in the Eq.(\ref{kn2}), we know
that the statistical-mechanical  entropy (\ref{kn2}) coincides
with the Mann-Solodukhin's result \cite{Mann96} which obtained by
using the conical singularity method.

\vspace*{0.4cm}
{\bf (B) The entropy of the Stationary
axisymmetric EMDA  black hole} \vspace*{0.4cm}

The stationary axisymmetric EMDA black hole metric  (we take the
solution b=0 in Eq.(14) in Ref.(2). The reason we use this
solution  is that the solution $b\neq0$ cannot be interpreted
properly as a black hole) is described by\cite{Garcia95}
\begin{eqnarray}
g_{tt}&=&-\frac{\bigtriangleup -a^2sin^2\theta }{\Sigma }, \ \ \ \
g_{t\varphi}=-\frac{asin^2\theta
[(r^2+a^2-2dr)-\bigtriangleup]}{\Sigma},\nonumber \\
g_{rr}&=&\frac{\Sigma}{\bigtriangleup}, \ \ \ \ g_{\theta \theta
}=\Sigma, \ \ \ \ g_{\varphi \varphi
}=\left(\frac{(r^2+a^2-2dr)^2-\bigtriangleup a^2sin^2\theta
}{\Sigma}\right)sin^2 \theta, \label{EMDA}
\end{eqnarray}
with
\begin{equation}
\Sigma=r^2-2dr+a^2cos^2\theta,\ \ \ \
\bigtriangleup=r^2-2mr+a^2=(r-r_+)(r-r_-),
\end{equation}
where $r_+=m+\sqrt{m^2-a^2}$, $r_-=m-\sqrt{m^2-a^2}$. The mass M,
the angular momentum J, the electric  charge Q,
 and the magnetic charge P  of the black hole are respectively given by
\begin{equation}
M=m-d,\ \ \ \ J=a(m-d),\ \ \ \ Q=\sqrt{2\omega d(d-m)}, \ \ \ \
P=g.
\end{equation}
By using the metric (\ref{EMDA}) we obtain
\begin{eqnarray}
& &\left\{\frac{\partial ^2g^{rr}}{\partial r^2}+
\frac{3}{2}\frac{\partial g^{rr}}{\partial r}\frac{\partial \ln
f}{\partial r}-\frac{2\pi}{\beta_H \sqrt{f}}\left(\frac{1}
{g_{\theta \theta}}\frac{\partial g_{\theta \theta}}{\partial r}+
\frac{1}{g_{\varphi \varphi}}\frac{\partial g_{\varphi
\varphi}}{\partial r}\right) -\frac{2g_{\varphi \varphi}}{f}
\left[\frac{\partial}{\partial r}\left(\frac{g_{t
\varphi}}{g_{\varphi
\varphi}}\right)\right]^2\right\}_{r_+}\nonumber
\\ &=&\frac{16r_+^2[(2d-r_+-r_-)r_+]+4d(8r_+-3d)
(r_+^2-2dr_++a^2)}{\Sigma^3}\nonumber \\ & &+
\frac{4[(3r_+-2d)(r_++r_--2d)-d^2]}{\Sigma^2}
+\frac{2d(r_++r_--2d)(\Sigma-2dr_+)}{\Sigma^3}\nonumber
\\ & & +\frac{2a^2(1+cos^2\theta )\Sigma-8a^2(r_+^2+a^2-2dr_+)cos^2\theta
}{\Sigma^3},\nonumber \\
 R&=&\frac{2a^2d^2sin^2\theta}{\Sigma^3}.
 \label{EMDA1}
\end{eqnarray}
Substituting Eq.(\ref{EMDA1}) into Eq.(\ref{smu}) and then taking
the integration of the $\theta$ and $\varphi$ we find  that the
statistical-mechanical entropy of the EMDA black hole is
\begin{eqnarray}
S&=&\frac{A_{\Sigma}}{48\pi}\sum^5_{i=0} \triangle _i
\mu_i^2ln\mu_i^2-\frac{1}{90}\left\{1+\frac{9d^2}{8r_+^2-16dr_+}
+\frac{9d\{3a^2d+(r_+^2-2dr_+)[\frac{4}{3}(r_++r_-)-d]\}}{16(r_+^2-2dr_+)^2}
\right. \nonumber \\ & & \left. \times
\left[1+\frac{r_+^2+a^2-2dr_+} {a\sqrt{r_+^2-2dr_+}}Arctan\left(
\frac{a}{\sqrt{r_+^2-2dr_+}}\right) \right] \right\}
\sum^5_{i=0}\triangle _iln\mu_i^2, \label{EMDA2}
\end{eqnarray}
where $A_{\Sigma}=4\pi(r_+^2+a^2-2dr_+)$. In order to compare the
entropy  (\ref{EMDA2}) with the thermodynamical entropy obtained
by the covariant Euclidean formulation, we \cite{Jing99}
calculated the thermodynamical entropy of the EMDA black hole  by
using the conical singularity method of the Ref.\cite{Mann96}. We
also find that the results obtained by the two methods are
equivalent.

From Eq.(\ref{kn2}) or Eq.(\ref{EMDA2}) we find the same result
for the Kerr black hole (setting $ Q=0$ in Eq. (\ref{kn2}) or
$d=0$ in Eq.(\ref{EMDA2})) as that Mann and Solodukhin found in
Ref.\cite{Mann96}, i. e., the quantum entropy  does not depend on
the rotation parameter $a$ and coincides with the quantum entropy
of the Schwarzschild black hole. We think the reason is that the
quantum entropy is mainly caused by quantum scalar fields near the
event horizon and in the region the scalar fields are co-rotating
with the black hole.

To summary,  by using BWM and with Pauli-Villars regularization
scheme, we investigate the statistical-mechanical entropy arising
from the minimally coupled quantum scalar fields rotating with the
angular velocity $\Omega_0$ in the general four-dimensional
non-extreme  stationary axisymmetric black hole space-time. An
expression of the statistical-mechanical  entropy is obtained for
the case $\Omega_0=\Omega_H$. The Kerr-Newman black hole and the
EMDA black hole are studied. It is shown that the
statistical-mechanical entropy obtained by using the formula
(\ref{smu}) and the quantum thermodynamical entropy derived from
the covariant Euclidean formulation (by using the conical
singularity method) are equivalent for the Kerr-Newman and the
EMDA black holes. The result fills in the gaps mentioned in
Ref.\cite{Frolov98} that the relation between the canonical and
covariant Euclidean formulations in the rotating black hole has
not been investigated. The study may provide us with a better
understanding of the relationship between the different entropies
to the black holes.

\end{document}